\documentclass[twocolumn,showpacs,preprintnumbers,amsmath,amssymb]{revtex4}
\usepackage{tabularx,graphicx}\begin{document}
\newcommand{\beq}{\begin{equation}}
\newcommand{\eeq}{\end{equation}}
\newcommand{\beqn}{\begin{eqnarray}}
\newcommand{\eeqn}{\end{eqnarray}}
\newcommand{\bmath}{\begin{subequations}}
\newcommand{\emath}{\end{subequations}}
\title{Why holes are not like electrons. II. The role of the electron-ion interaction}
\author{J. E. Hirsch }
\address{Department of Physics, University of California, San Diego\\
La Jolla, CA 92093-0319}
 
\date{September 1, 2004} 
\begin{abstract} 
In recent work, we discussed the difference between electrons and holes in energy band
in solids from a many-particle point of view, originating in  the electron-electron interaction,
and argued that it has fundamental consequences for superconductivity. 
Here we discuss the fact that  there is also a   fundamental
difference between electrons and holes already at the single particle level, arising from the 
electron-ion interaction. The difference between electrons and holes due to this effect parallels the difference
due to electron-electron interactions: {\it holes are more dressed than electrons}. We propose that superconductivity
originates in 'undressing' of carriers from $both$ electron-electron and electron-ion interactions, and that
both aspects of undressing have observable consequences.
\end{abstract}
\maketitle 
\section{Introduction}

Hamiltonians used to describe many-body phenomena in solids are usually electron-hole symmetric (by 
electrons and holes it is  meant the charge carriers at the Fermi energy when the Fermi level is near
the bottom and near the top of the band respectively). Instead, in the first paper of this series\cite{hole2} and other 
recent work\cite{hole2p} we have argued that holes
are fundamentally different from electrons, due to the different effect of electron-electron interaction for
carriers at the bottom and the top of a band. We have proposed a new class of model Hamiltonians, 
'dynamic Hubbard models', to
describe this physics, and argued that this physics plays a fundamental role in 
superconductivity\cite{dynh,dynh2}. 
These electron-hole $asymmetric$ Hamiltonians describe quasiparticles that become increasingly dressed by the
electron-electron interaction as the Fermi level rises from the bottom to the top of the band, and  give
rise to superconductivity driven by quasiparticle 'undressing'\cite{undr}. They also display
 many characteristic features that are observed
in high $T_c$ cuprates.  

In these Hamiltonians, the electron-electron interaction breaks electron-hole symmetry, however the
single particle part of the Hamiltonians is still electron-hole symmetric. In this paper   we point out that 
in real solids a
fundamental electron-hole asymmetry already exists at the single-electron level, which parallels the 
electron-hole asymmetry arising from electron-electron interactions. This physical effect is $also$ missing in the
 tight binding Hamiltonians commonly used  to describe correlated electrons in solids. Just as quasiparticles 
in real solids are increasingly
dressed by {\it electron-electron} interactions as the Fermi level rises in the band\cite{hole2}, we point out here that they are
also increasingly dressed by the {\it electron-ion} interaction as the Fermi level rises. Furthermore we argue that
because when holes pair the band becomes locally less full, 'undressing'  from $both$ the electron-electron
interaction $and$ the electron-ion interaction will take place. Remarkably, as we will discuss, experimental evidence that
'undressing' from the electron-ion interaction takes place upon the transition to the superconducting state has
been known, even if not fully appreciated, for a long time. The fact that electrons
'undress' from the electron-ion interaction when they pair has also fundamental consequences 
for superconductivity\cite{atom}.

The central character in this paper is in fact not the hole, but the 'invisible' antibonding electron, the electron at the Fermi level
when the Fermi level is near the top of the band. Using the
language of 'holes' rather than 'electrons' in fact obscures the essential physics since  these  antibonding electrons are the
ones  that   'undress' and
   carry the supercurrent (as electrons, not as holes) in the superconducting state. 

\section{Tight binding electronic energy bands}
The tight binding Hamiltonians usually considered such as the Hubbard model have a single-particle part of the form
\beq
H_0=-\sum_{ij\sigma}t_{ij} c_{i\sigma}^\dagger c_{j\sigma}
\eeq
where $c_{i\sigma}^\dagger$ creates an electron in Wannier orbital $\varphi_i(r)$ centered at lattice site $i$. This
Hamiltonian may or may not be electron-hole symmetric. However  if we  restrict ourselves to nearest neighbor hopping
on a hypercubic lattice as is usually done:
\beq
H_0=-t\sum_{<ij>\sigma}( c_{i\sigma}^\dagger c_{j\sigma} + h.c.)
\eeq
then the Hamiltonian is electron-hole symmetric, as can be seen from the fact that the canonical transformation
\beq
c_{i\sigma}=(-1)^i d_{i\sigma}^\dagger
\eeq
leaves it invariant. The band energy is
\beq
\epsilon_k=-2t\sum_{\nu=1}^d cos k_\nu a 
\eeq
with $a$ the lattice spacing and $d$ the dimensionality.
The effective mass for carriers at the bottom of the band is independent of direction and given by
\beq
m^*=[\frac{1}{\hbar^2}\frac{\partial^2 \epsilon_k}{\partial k_x^2})_{k=0}]^{-1}=\frac{\hbar^2}{2 t a^2}
\eeq
and the effective mass at the top of the band has the same magnitude as Eq. (5)
and opposite sign.

When one includes electron-electron interactions that don't break electron-hole symmetry, e.g. in an extended 
Hubbard model
\beq
H=H_0+\sum_{ij} V_{ij}n_i n_j
\eeq
or electron-phonon interactions as in the Holstein or Su-Schrieffer-Heeger models, or electron-spin interactions as in the
Kondo lattice model, the Hamiltonian retains electron-hole symmetry and hence the properties of the system
are identical for band filling $n_e$ and $2-n_e$, with $n_e$ the number of electrons per site. Instead, adding a
correlated hopping term\cite{corrhop} or electron-boson interactions of particular forms as in 
dynamic Hubbard models\cite{dynh,dynh2}
breaks electron-hole symmetry and leads to qualitatively new physics.

\begin{figure}
\includegraphics[height=.20\textheight]{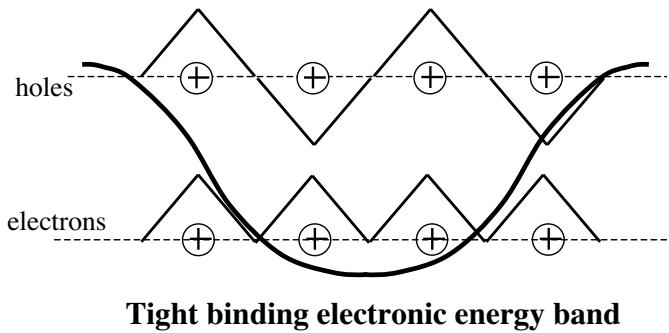}
  \caption{Electronic states in a tight binding band. The states   at the top of the band and those at the
bottom of the band are related by a canonical transformation. This will be the case if they are
Bloch sums  of $non-overlapping$ Wannier orbitals  with wavevector $k$ and $\pi-k$ respectively.
  }
\end{figure}

Here however we want to focus on the fact that in writing down the non-interacting Hamiltonian Eq. (2)
we have already eliminated an important source of electron-hole asymmetry, arising from the
electron-ion interaction. The electron-hole transformation Eq. (3) says that the wavefunction of an electron at the
bottom of the band is essentially the same (except for a phase factor) as the one for an electron at the top
of the band. Note also that the Wannier orbitals used to define the Hamiltonian Eq. (2) need to be
orthogonal to each other. Schematically, this can be represented by the wavefunctions shown in Figure 1.
Here indeed the states at the bottom and the top of the band are equivalent, since one can be transformed
  into the other by the operation $\varphi_i(r)\rightarrow (-1)^i\varphi_i(r)$. However in fact 
   the real situation is very different and no such electron-hole symmetry exists even
at the single particle level.
  
\section{Real electronic energy bands}
Real electronic energy bands can be obtained from band structure calculation schemes, and the eigenstates certainly don't look
like the states depicted in Figure 1. 
Here we wish to focus on what we believe is the key universal aspect that differentiates the states at the bottom
and the top of any electronic band.

The essential physics can be illustrated clearly with a diatomic molecule. If $\varphi_i(r)$ is an atomic orbital at atom $i=1,2$, the
bonding and antibonding atomic orbitals in an LCAO scheme are given by
\bmath
\beq
\phi_{b,a}=\frac{\varphi_1(r) \pm \sigma \varphi_2(r) }{[2(1\pm \sigma S)]^{1/2}}
\eeq
\beq
S=(\varphi_1,\varphi_2)
\eeq
\beq
\sigma=S/|S|
\eeq
\emath
with the upper (lower) sign corresponding to bonding (antibonding) states. For $s-$orbitals, the sign of the overlap matrix element is positive and the $even$ linear combination gives the 
lowest energy molecular orbital, the bonding orbital, and the odd linear combination gives the high energy
antibonding orbital. For $p-$orbitals $S$ is negative and the situation is reversed. However the key point is  that the lowest molecular orbital is always 
the linear combination that yields high electronic charge density between
the ions, and the other one has a node in the electronic wavefunction at the midpoint between the ions.
Conversely, the low energy bonding orbital has always lower charge density at the ion site than the 
antibonding orbital.

For an energy band we argue that the electronic states look qualitatively as shown in Figure 2, independent
of which atomic orbital gave rise to the band. The point is that  the states at the bottom of the band have lowest
energy (by definition), which is achieved by piling up electronic charge density in the region where it can
most benefit from the electron-ion potential, namely in the interstitial region between ions; at the same time, due 
to normalization, the charge density at the ionic site is reduced compared to the free atom situation, and the
resulting smooth wave function also has a low kinetic energy. Instead, the states at the top of the band
are constrained by the fact that they have to be orthogonal to the states below them. This causes the wavefunction to
 have a node   in the region between the ions, and a higher amplitude at the ionic
site than for the isolated ion; as a consequence, these states have a high potential energy, since they don't 
take maximal advantage of the electron-ion potential, and a high kinetic energy because the
wavefunction is 'spiky' rather than smooth. 

Because the wavefunction at the bottom of the band is
more smooth, it resembles more the free electron plane-wave function, which gives rise to a uniform
electronic density. The wavefunctions at the top of the band give rise to a non-uniform charge density, quite unlike
free electron wave functions . Furthermore the effective mass
defined by Eq. (5) will be positive near the bottom of the band, as the free electron mass, and negative
near the top of the band. We conclude that quite generally the single-particle electronic states near the
bottom of the band are more free-electron-like than those near the top of the band.

\begin{figure}
\includegraphics[height=.20\textheight]{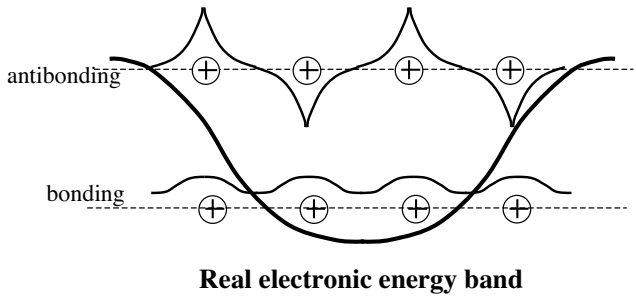}
  \caption{Electronic states in a real band. The states at the bottom (bonding) have a high density of charge 
in-between the ions, and a  smooth wave function. The states at the top (antibonding)
 have a node in the charge density between the ions and a spiky wave function.}
\end{figure}

We argue that these are universal physical differences between electronic states at the bottom and top of
electronic energy bands, determined by the  facts that states at the bottom have low energy and states
at the top have high energy, and by the exclusion principle. These are real physical differences that 
cannot be eliminated away by canonical transformations. For $s-$orbitals the energy versus $k$ relation
looks like that in Figure 2, while for $p-$orbitals it is inverted, with the lowest states at $k\sim \pi $ and the
highest states at $k\sim 0$. In both cases the electronic wavefunction looks qualitatively as in
Figure 2 in the region between the ions, except that for a band deriving from $p-$orbitals the
wavefunction has an additional node exactly at the ionic site for all states in the band.
The same considerations apply to bands originating in other orbitals.

The physical difference between states at the bottom and top of the band is also embodied in their name,
bonding and antibonding respectively. The high interstitial electronic density of the bonding orbitals
gives rise to an attractive interaction between ions, binding the solid together; instead, the vanishing
electronic charge density between ions of the antibonding electrons causes a repulsive interaction between
ions, which tends to break the solid apart. This is why lattice instabilities are associated with the 
presence of antibonding states at the Fermi energy, i.e. with bands that are almost full. As is well known,
superconductivity is also often associated with the presence of lattice instabilities nearby in the phase
diagram\cite{matthias}, indicating a connection between antibonding states and superconductivity.

\section{Weak coupling}

It was noted already by Bloch that the very different tight binding and weak binding starting points for the description of electronic
states in solids  give complementary and very 
similar pictures. The fact that electrons at the bottom of the band are more similar to free electrons than those at the
top of the band discussed in the last section is very evident   from the weak coupling point of view. Perturbation theory in the
electron-ion potential $U_K$ yields for the band energy
\beq
\epsilon_k=\epsilon_k^0+\sum_K \frac{|U_K|^2}{\epsilon_k^0-\epsilon_{k-K}^0}
\eeq
with $\epsilon_k^0=\hbar^2k^2/2m_e$ the free electron energy ($m_e=$free electron mass) and $K$ reciprocal
lattice vectors. Starting from the state at $k=0$ energy denominators are large, so the
second term in Eq. (8) is small and the energy versus $k$ relation is almost free-electron-like.
The wavelength $\lambda=2\pi/k$ is large and the electronic wave function
\bmath
\beq
\varphi_k=\varphi_k^0+\sum_K\frac{U_K}{\epsilon_k^0-\epsilon_{k-K}^0} \varphi_{k-K}^0
\eeq
is almost the free electron plane wave
\beq
\varphi_k^0=\frac{1}{\sqrt{V}}e^{1\vec{k}\cdot\vec{r}}
\eeq
\emath
The effective mass tensor is given by 
\beq
(\frac{1}{m^*})_{ij}=\frac{1}{\hbar^2}\frac{\partial^2\epsilon_k}{\partial k_i \partial k_j}=
\frac{1}{m_e}\delta_{ij}+\frac{1}{\hbar^2} \frac{\partial^2}{\partial k_i \partial k_j}
\sum_K \frac{|U_K|^2}{\epsilon_k^0-\epsilon_{k-K}^0}
\eeq
and is almost isotropic and free-electron-like for small $k$ since the second term in
Eq. (10) is small.

As the Fermi level rises electrons at the Fermi surface become increasingly
'dressed' by the electron-ion interaction: the wavevector $k$ increases, the
energy denominators in Eqs. (8)-(10) decrease and the electronic energy,
wavefunction and effective mass increasingly deviate from the free electron values,
as the second terms in Eqs. (8)-(10) become increasingly important compared to the
first terms. Qualitatively, as the wavelength $\lambda$ decreases, the electrons become
increasingly 'aware' of the existence of the discrete ionic potential due to the fact that the wavelength of
the electronic wavefunction becomes closer to that of the ionic potential. 

 Consider for definiteness a simple cubic lattice. The perturbative expressions
 Eqs (8)-(10) break down when the wavevector $k$ approaches one half of the 
 smallest reciprocal lattice vectors $\vec{K}=\frac{2\pi}{a} \hat{l}$, with
 $\hat{l}$ one of the three principal axis, i.e. when the wavevector approaches the edge
 of the Brillouin zone, or the electronic wavelength approaches twice the ionic charge wavelength.  Then, nearly degenerate perturbation theory yields for the
 state near the top of the band
 \bmath
 \beq
 \epsilon_k=\frac{\epsilon_k^0+\epsilon_{k-K}^0}{2}-
 \sqrt{(\frac{\epsilon_k^0-\epsilon_{k-K}^0}{2})^2+|U_K|^2}
 \eeq
 \beq
 \varphi_k=\frac{\varphi_k^0+\varphi_{k-K}^0}{\sqrt{2}}
 \eeq
 \beq
 \frac{1}{m^*}=\frac{1}{m_e}-\frac{\hbar^2}{4m_e^2}\frac{K^2}{|U_K|}
 \eeq
 \emath
 The wavefunction  
 $\varphi_k(r)\sim cos(kr), k\sim \pi/a$   is zero at the
 midpoint between the atoms, just as the tight binding picture also predicts. Because the ionic potential has broken the degeneracy
 with the other linear combination of free electron wave functions, $\varphi'_k(r)\sim sin(kr)$, the charge density associated
 with the states at the top of the band is non-uniform unlike the free electron case.
 Beyond lowest order perturbation theory these wavefunctions get modified by the ionic potential that pulls the electronic charge
 closer to the regions of positive charge.
 The dependence of 
 energy on wavevector is very different from the free electron case, and in particular the 
 effective mass Eq. (11c) is negative for small $|U_K|$ which is the regime where these expressions
 are valid.
 
 These facts are of course well known\cite{kittel}. However it is not usually stressed that they reflect
 a fundamental physical difference between states at the bottom and the top of electronic energy bands,
 i.e. between bonding and antibonding electrons, or equivalently between electrons and holes.

 For the lowest band in a solid these considerations then imply that electronic states at the
 bottom of the band are not very different from free electron states, and in particular that
 the electronic density is nearly uniform and not strongly modulated by the ionic potential.
 The electron-ion interaction has an increasing effect in modifying the electronic density
 and the energy-momentum relation from the free electron values as the states approach
 the top of the band. We argue that the same physics will be true for other bands.
 Let us first consider the results of the perturbation theory discussed above for the second band. The
 wavefunction $\varphi'_k(r)\sim sin(kr)$ at the bottom of the second band has a node at the ionic
 site but is smooth in the interstitial region, as free electron wave funcions are. The effective mass at the
 bottom of the second band is given by Eq. (11c) with a positive sign for the second term, hence is positive as for
 free electrons. Even though it is smaller than the free electron mass, second order contributions of the form Eq. (8) 
 will increase its value towards the free electron mass. For example, the dominant contribution in second order
 for $k\sim \pi/a$ comes from $K'=4\pi/a$ in Eq. (8) and yields
 \beq
 \frac{1}{m^*}=\frac{1}{m_e}[1+\frac{\hbar^2\pi^2}{m_e a^2 |U_K|}-\frac{a^4 m_e^2 |U_{K'}|^2}{2\pi^4 \hbar^4}]
 \eeq
and contributions from other $K$ values will increase it further
 . At the top of the
 second band i.e. at $k=2\pi/a$, degenerate perturbation theory with $K=4\pi/a$ again yields negative effective mass.
 More generally,   we know from pseudopotential theory\cite{harrison} that one can find an effective
 description of conduction  bands in solids that is similar to the lowest band discussed above. In pseudopotential 
 theory the conduction band energy is to second order  
 \beqn
\epsilon_k&=&\epsilon_k^0+<k|W|k>+  \nonumber \\
  & &\sum_K \frac{<k|W|k-K><k-K|W|k>}{\epsilon_k^0-\epsilon_{k-K}^0}
\eeqn
where the pseudopotential $W$ is an operator rather than a local function of position. The pseudopotential is chosen to 
give a smooth 'pseudo wave function' to optimize the convergence, and it is found that its matrix elements are small and
the second order expression Eq. (13) is adequate except near Bragg planes, as in the simple case discussed above\cite{harrison}. 
Eq. (13) yields an additional contribution to the effective mass from the first order perturbation term that is absent
in Eq. (8), however that term is generally found to be small\cite{harrison}. Consequently the considerations made above for the lowest band
still apply. The pseudowavefunctions (linear combination of orthogonalized plane waves) give a smooth charge distribution in the interstitial regions
near the bottom of the band\cite{harrison}, as free electrons do. Finally, quite generally it is   true that the effective mass is positive, hence closer to
 the free electron value, near the bottom of a band, and negative, hence more
 different from the free electron value, near the top of a band.

\section{Physical differences between electrons and holes due to the electron-ion interaction}

In the previous sections we have discussed the difference in wavefunction, resulting charge density,
 and energy-wavevector relation
between states at the bottom of a band (electrons) and states at the top of a band (holes), and argued that
states near the bottom of the band are free-electron-like and those near the top of the band are not. A tight binding model
of the form Eq. (2) does not reflect the physical difference between bonding and antibonding electrons. Yet these differences have
concrete observable consequences, as discussed in what follows.

\subsection{Momentum transfer to the lattice}
When a force is applied to an electron inside a metal, both the electron and the lattice change their momentum:
\beq
\vec{F}=\frac{\Delta \vec{p}}{\Delta t}=\frac{\Delta \vec{p}_{el}+ \Delta \vec{p}_{latt}}{\Delta t}
\eeq
Semiclassical transport theory relates the change in total momentum to the change in the electronic
$crystal$ $momentum$
\beq
\Delta \vec{p}=\hbar\Delta \vec{k}
\eeq
while the change in electronic momentum is given by
\beq
\Delta \vec{p}_{el}=m_e\Delta \vec{v}_{el}=m_e\frac{1}{\hbar^2}\frac{\partial^2\epsilon_k}{\partial k^2}
\hbar\Delta\vec{k}=\frac{m_e}{m^*} \Delta \vec{p}
\eeq
(assuming isotropic effective mass for simplicity) so that the momentum transfered to the lattice is
\beq
\Delta \vec{p}_{latt}=(1-\frac{m_e}{m^*})\Delta \vec{p}
\eeq
For electrons near the bottom of the band, $m^*$ is close to $m_e$ and practically all the momentum is
transfered to the electron and none to the lattice. Instead, for electrons near the top of the band the
change in the electron momentum is opposite to the transfered momentum since $m^*$ is negative,
and the lattice needs to pick up both the externally transfered momentum and the negative of the
momentum change of the electron. We may quantify the 'dressing' of the free electron by the
momentum transfered to the lattice when an external force attempts to change the electronic
momentum;  it is clear then that the electron-ion interaction increasingly 'dresses' the bare electron at
the Fermi level as the Fermi level rises in the band.

\subsection{Conduction of electricity}

When an electric potential difference is applied to a metal, electric current flows from the higher to the lower
potential side of the metal. However, the behavior is very different for electrons near the bottom and near
the top of the band. The change in velocity of an electron upon application of an electric field $\vec{E}$ is
\beq
\Delta\vec{v}=\frac{1}{m^*}e\vec{E}\tau
\eeq
with $\tau$ the collision time. Electrons in the lower half of the band have $m^*>0$ and hence change their velocity
in the direction that contributes to the flow of electricity (i.e. opposite to $\vec{E}$ since $e<0$); instead,
electrons in the upper half of the band change their velocity in direction that opposes the flow of electric current,
and as the band becomes filled the two contributions exactly cancel and zero current results. Hence
the 'dressing' of the antibonding electron by the electron-ion potential causes it to oppose, rather than
contribute, to the conduction of electricity as a free electron would.

\subsection{Optical conductivity}

The integrated optical conductivity from $intra-band$ transitions, when the Fermi level is close to the bottom
of the band, is given by 
\beq
\frac{2}{\pi  e^2}\int_{intraband} d\omega\sigma_1(\omega)=\frac{n_e}{m^*}
\eeq
with $n_e$ the number of electrons in the band. Hence electrons near the bottom of the band each contribute
a positive amount to the low frequency 
optical conductivity, which is close to the contribution of a free electron if $m^*$ is close
to $m_e$. Instead, when the Fermi level is close to the top of the band the integrated intra-band optical conductivity is
\beq
\frac{2}{\pi  e^2}\int_{intraband} d\omega\sigma_1(\omega)=\frac{n_h}{|m^*|}=\frac{2-n_e}{|m^*|}
\eeq
and each antibonding electron added to the nearly full band $decreases$ rather than increases the intra-band optical
conductivity. The difference between the Drude weight Eq. (19) and the Drude weight that would arise from
free electrons, $n_e/m_e$, also quantifies the amount of 'dressing', and this difference increases as the
Fermi level rises from the bottom to the top of the band. The global conductivity sum rule
\beq
\frac{2}{\pi  e^2}\int_{0} ^\infty d\omega\sigma_1(\omega)=\frac{n_e}{m_e}
\eeq
implies that this 'missing' spectral weight is transfered from low intra-band frequencies to high inter-band frequencies due to the
electron-ion interaction.

\subsection{Hall effect}

As a final physical manifestation of the difference between electrons at the bottom and at the top of electronic energy bands we mention
the Hall effect\cite{peierls}. Electrons near the bottom of the band respond to crossed electric and magnetic fields as free
electrons would, namely they traverse cyclotron orbits in the direction consistent with the negative charge
of the free electron. Instead, the strong 'dressing' of the free electron by the electron-ion potential for electrons
near the top of the band causes them to respond as if they had a charge of opposite sign, reflecting the
positive charge of the ionic lattice that dresses them.

\subsection{Summary}

In summary, we have discussed in this section various observable 
manifestations of the physical difference between electrons at the bottom
and at the top of electronic energy bands, which arise due to the electron-ion interaction. Electrons near the bottom of
the band resemble free electrons with a nearly uniform charge density, and are largely unaffected by the
presence of the discrete ionic lattice potential. When perturbed by external probes they respond very much 
like free electrons. Instead, electrons near the top of energy bands (antibonding electrons) have a wave function that changes rapidly
over interatomic distances as shown in Fig. 2,  are tightly coupled to the discrete ionic lattice, and their charge density is very
non-uniform and hence different from the free electron case. The antibonding electrons are 
strongly 'dressed' by the electron-ion interaction. When
perturbed by external probes, this tight coupling between antibonding electrons and positive ionic charge causes them to
respond very differently from free electrons.

\section{Dressing from electron-electron interactions}

In previous work we have discussed the different effect of the electron-electron interaction for electrons at the bottom and the top of bands\cite{hole2,hole2p,dynh}. Just as the electron-ion interaction discussed in the previous sections, we showed that the electron-electron
interaction increasingly dresses the quasiparticle as the Fermi level goes up in the band. Dressing due to the
electron-electron interaction does not change the sign of the effective mass but increases its magnitude. It also 
causes another effect that goes beyond single-particle physics, it reduces the quasiparticle weight in the single particle
spectral function and gives rise to incoherent spectral weight at higher energies. For the optical conductivity,
$both$ the dressing from electron-ion and from electron-electron interaction cause spectral weight to be pushed up
from the low-frequency intra-band range to higher frequencies.

In summary, both the electron-ion and the electron-electron interaction cause electrons in a metal to become 'dressed',
i.e. different from free electrons. The dressing from both of these sources becomes increasingly important as the
Fermi level goes up in the band, and is largest when the Fermi level is close to the top of the band.
We adopt then as a basic principle: {\it higher concentration of electrons in a band leads to higher
dressing of the quasiparticles at the Fermi energy. }

\section{Superconductivity from undressing}

When the Fermi level is close to the top of the band, the carriers at the Fermi energy, antibonding electrons,
 are most highly dressed.
Furthermore, the kinetic energy of electrons at the Fermi energy is highest, and they do not benefit from the 
crystal ionic potential because their charge density in the region between ions is small. If these electrons   were
able to occupy states that are lower in the band their  energy would be lowered. However, those lower states
are occupied by other electrons (bonding electrons) and the Pauli principle prevents other fermions from
occupying those states.  In the absence of electron-electron interactions, the antibonding electrons have no
choice but to remain in the 'unfavorable' antibonding states, and to pay the high price in both kinetic and
potential energy in doing so.

However, electrons do interact with each other, and states at the Fermi energy can be modified by electronic
correlations. When the Fermi level is close to the top of the band, pairing of holes leads locally to a higher hole
concentration, hence to a lower electron concentration. According to the basic principle enunciated above,
this should lead to 'undressing' of carriers at the Fermi energy. Since both the dressing by electron-electron and by
electron-ion interactions increase with band filling, a local decrease in band filling should lead to 
'undressing' from both the electron-electron and the electron-ion interactions, as shown schematically in 
Figure 3. A Cooper pair behaves as a boson rather than a fermion, and these arguments indicate that the
members of a Cooper pair will bear a closer resemblance to free undressed electrons than the unpaired antibonding electrons.

\begin{figure}
\includegraphics[height=.10\textheight]{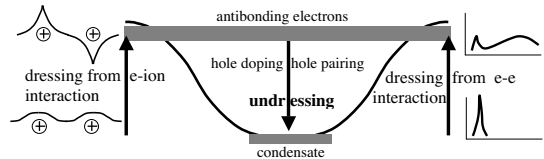}
  \caption{As the Fermi level rises in a band,   electrons at the Fermi energy 
 become  dressed due to electron-electron interactions which modify the free particle
  spectral function (as shown schematically on the right of the figure), and due to electron-ion interactions which modify the
  free-electron wave function (as shown on the left side of the figure). Pairing effectively lowers the position
  of the Fermi level and causes the carriers at the Fermi energy to 'undress' and become
  free-electron-like.}
\end{figure}

The phenomenology of undressing from the electron-electron interaction is described mathematically by dynamic
Hubbard models\cite{dynh} and by their low energy effective counterpart, the Hubbard model with
correlated hopping\cite{corrhop}. The quasiparticle weight is given by\cite{undr}
\beq
z(n_e)=[1-(1-S)\frac{n_e}{2}]^2
\eeq
with $0\leq n_e\leq 2$ and $S<1$ a parameter that depends on the nature of the ion\cite{hole2}.
The effective mass is given by $m^*= m_e/z(n_e)$. As the band filling $n_e$ decreases,
$z(n_e)$ increases and $m^*$ decreases.

Experimentally, undressing from electron-electron interaction is seen as an increase in the coherent
response in photoemission experiments\cite{ding}, reflecting increase in the quasiparticle weight,
and transfer of optical spectral weight from high to low frequencies, reflecting decrease in the
quasiparticle mass and decrease in the kinetic energy\cite{marel}. 
In the following we discuss experimental evidence for undressing from the
electron-ion interaction upon onset of superconductivity.

\section{Undressing from the electron-ion interaction}

There are several experiments that indicate that in the transition to superconductivity 'undressing' from the
electron-ion interaction also takes place:

\noindent {\it (1) Hall effect}: As discussed earlier, the Hall coefficient is negative for electrons near the
bottom of the band that are 'undressed' from the electron-ion interaction and is positive for electrons near
the top of the band that are dressed by the electron-ion interaction. Hence 'undressing' from the electron-ion
interaction should be signaled by a change in sign of the Hall coefficient from positive to negative. Indeed,
empirical evidence shows that superconductivity is associated with a positive Hall coefficient in the normal state
in the vast majority of cases\cite{kikoin,chapnik}, indicating that the carriers at the Fermi energy are dressed antibonding electrons.
Furthermore, it is found in both high $T_c$\cite{sign} and low $T_c$  materials\cite{sign2} that the 
Hall coefficient changes its sign from positive to negative at 
temperatures slightly below $T_c$, which indicates that carriers change from hole-like to electron-like.

\noindent {\it (2) Bernoulli potential}:
Because the superfluid carriers in superconductors carry kinetic energy one expects that an electric field should exist
in regions where there is a spatial variation of the superfluid velocity, according to the relation
\beq
\vec{E}=\frac{1}{e} \vec{\nabla}\frac{1}{2} m_e v_s^2  .
\eeq 
This was first discussed by London for a free electron model\cite{london}, and the resulting potential is termed 'Bernoulli potential'.
As discussed by Adkins and Waldram\cite{adkins}, within BCS theory the sign of the effect should correspond to the sign of the
charge carriers in the normal state. Experiments to measure this effect have been performed with
samples of $Pb$, $Nb$, $PbIn$ and $Ta$, and in all cases the sign of the effect measured corresponds to the superfluid
carriers having $negative$ $charge$\cite{bok,morris,chiang}. Note that the Hall coefficient in the normal state is $positive$ for all these cases.
Furthermore the magnitude of the effect measured is consistent with the mass in Eq. (23) being the {\it free electron mass}. \cite{chiang}

\noindent {\it (3) Rotating superconductor}:
A superconducting body rotating with angular velocity $\vec{\omega}$ develops a uniform magnetic field in its interior\cite{becker}.
This can be understood as follows: as the ions start rotating, a time-dependent magnetic field is generated which in turn
induces an azimuthal electric field according to Faraday's law
\beq
\oint \vec{E}\cdot \vec{dl}=-\frac{1}{c} \frac{d}{dt}\int \vec{B}\cdot\vec{ds}
\eeq
so that if the magnetic field is uniform the electric field at position $\vec{r}$ from the axis of rotation is
\beq
\vec{E}=\frac{1}{2c}\vec{r}\times \frac{d \vec{B}}{dt}
\eeq
Now semiclassical transport theory relates the electric field to the time derivative of the wavevector of the carrier
\beq
\hbar\frac{d \vec{k} }{dt}=e\vec{E}
\eeq
while the time derivative of the velocity of the carrier is given by (for an isotropic case)
\beq
\frac{d\vec{v}}{dt}=\frac{d}{dt}(\frac{1}{\hbar}\frac{\partial \epsilon}{\partial \vec{k}})=\frac{1}{m^*}\hbar\frac{d \vec{k} }{dt}
\eeq
so that
\beq
\frac{d {\vec{v}}}{dt}=\frac{e}{m^*}\vec{E}=\frac{e}{2m^*c}\vec{r}\times \frac{d\vec{B}}{dt}
\eeq
In steady state  the superfluid in the interior rotates together with the lattice\cite{becker} so that $\vec{v}=\vec{\omega}\times\vec{r}$
  and from integration of Eq. (28)
\beq
\vec{\omega}=-\frac{e}{2m^*c}\vec{B} \eeq
If the superfluid carriers were 'dressed' by the electron-ion interaction
the effective mass in Eq. (29) would be dependent on the particular material
and in particular    $\vec{B}$ would point antiparallel to $\vec{\omega}$ if the carriers are hole-like. Instead, it is found 
experimentally that for all superconductors where it has been measured (including
high $T_c$ cuprates and heavy fermion materials) \cite{hildebrand} 
\beq
\vec{B}=-\frac{2m_ec}{e}\vec{\omega} .
\eeq
with $m_e$ the {\it free electron mass}.
 The fact that the magnetic field always points 
parallel and never antiparallel to the angular velocity indicates that the superfluid carriers have $negative$ charge.
The fact that the magnitude of the magnetic field is given by Eq. (30) for all materials, with $m_e$ the $bare$ electron mass,
indicates that the carriers in the superconducting state are   undressed free electrons .
This means that the dressed carriers at the top of the Fermi distribution in the band depicted in Fig. 3, with antibonding wave function that knows about the discrete ionic potential,
 condense to the bottom of the Fermi distribution with a smooth long wavelength wavefunction that is insensitive to the short wavelength 
 ionic potential.
 Physically the magnetic field Eq. (30) arises because the negative electrons near the surface of the superconductor
lag behind and rotate at slightly smaller angular velocity than the body, as shown schematically in Fig. 4.

\begin{figure}
\resizebox{6cm}{!}{\includegraphics[width=9cm]{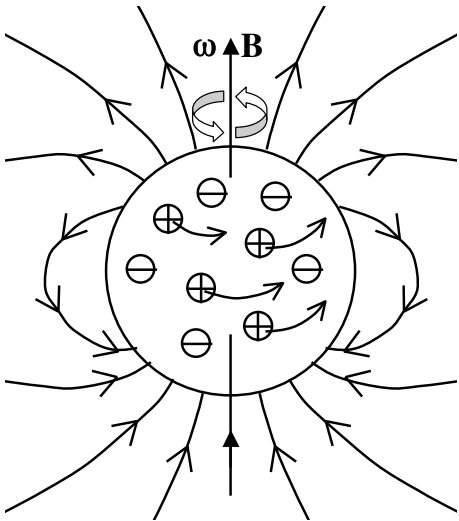}}
\caption{Experimental proof that electrons undress from the electron-ion interaction when they become superconducting.
The $magnitude$ of the magnetic field generated in a rotating superconductor Eq. (30) 
reflects relative motion of superfluid electrons with $bare$ mass $m_e$ .
The $sign$ of the magnetic field generated reflects slowing down of superfluid $negative$ charge.
}
\label{fig4}
\end{figure}

\noindent {\it (4) Gyromagnetic effect}:
A related effect occurs if a magnetic field is suddenly applied to a superconductor at rest. The supercurrent that
develops to nullify the magnetic field in the interior (Meissner effect) carries angular momentum, and for the
total angular momentum of the superconducting body to be unchanged the body has to start rotating with
angular momentum in the opposite direction. If the superfluid carriers have negative charge, the body will start 
rotating with angular velocity antiparallel to the applied field, which is indeed what is observed\cite{gyro}.

\section{Kinetic energy lowering, wavelength expansion and phase coherence}
An effect associated with having an increasing number of electrons in a band is of course an increase in the kinetic energy of the
electrons at the Fermi energy. This is true both for band electrons as well as for free electrons. The kinetic energy associated
with a spatial variation of the wavefunction in a region of linear dimension   $\lambda$ is
\beq 
T\sim \frac{\hbar^2}{2m_e \lambda^2}
\eeq  $\lambda$ can be thought of as the wavelength of the  electronic wavefunction in a $k$-space picture, or equivalently as the linear
dimension of the region occupied by each electron, i.e. the mean distance between electrons, in a real space picture.
In a free electron model, $\lambda \sim k_F^{-1}$. 
 For a single electron in an empty band $\lambda$ in Eq. (31) is the linear dimension of the sample.
 As more electrons are added to a band the wavelength of the  electronic wave function decreases, or equivalently
 the size of the region occupied by each electron decreases and the wavefunction becomes more spatially confined,
which leads to an increase of kinetic energy.   To the extent that superconductivity is
associated with kinetic energy lowering\cite{marel,science}
it is natural to expect that this will occur when the electrons at the Fermi level have highest kinetic energy in the
normal state, which corresponds to the case of an almost full band, which also corresponds to the  smallest spatial extent of the
electronic wavefunction, with $\lambda$ of order the interatomic
spacing $a$. If the kinetic energy Eq. (31) decreases as the system goes superconducting
  it implies that the wavelength of the electronic wavefunction increases so that it no longer 'sees' the short 
  wavelength ionic potential and becomes free-electron-like, and $\lambda$ in Eq. (31) becomes again the linear macroscopic
  dimension of the sample, as for the empty band. 
  
  We can also understand the origin of diamagnetism in superconductors from this point of view. The diamagnetic response of
  a normal metal (Landau diamagnetism) can be understood as arising from induced Ampere circular currents
  of radius given by $r \sim \lambda \sim k_F^{-1}$, the interelectronic spacing or equivalently the wavelength of the electronic
  wavefunction. In a free electron model
  \beq \chi_{Landau}=-\frac{1}{3}\chi_{Pauli}=-\frac{1}{3}\mu_B^2g(\epsilon_F)
  \eeq
  with $\mu_B=e\hbar/2m_ec$ the Bohr magneton and $g(\epsilon_F)=3n/2\epsilon_F$ the density of states, with $n$ the number
  of electrons per unit volume. The Larmor diamagnetic response from $n$ 'atoms' per unit volume is
  \beq
  \chi_{Larmor}=-\frac{e^2n}{6m_e c^2 } <r^2>
  \eeq
  with $<r^2>$ the spatial extent of the electronic wavefunction. Eqs. (32) and (33) are the same for $<r^2>=3/(2k_F^2)\sim\lambda^2$. When the metal 
  goes superconducting the Meissner currents extend over the entire sample, and the wavelength
  $\lambda$ becomes $R$,  the macroscopic dimension of the sample. Using Eq. (33) for the atomic susceptibility
  yields perfect diamagnetism when $<r^2>$ becomes macroscopic\cite{gins}. Hence we can interpret the change from 
  Landau diamagnetic response to London diamagnetic response as a wavelength expansion from
  $\lambda \sim k_F^{-1}\sim a$ ($a=$lattice spacing) for the electrons at the top of the Fermi distribution in
  Fig. 3 to $\lambda\sim R\sim k^{-1}$, i.e. the bottom of the band in Fig. 3 where the free-electron states are.
  
  Furthermore the concept of  'wavelength expansion' provides a qualitative understanding of the phase coherence in superconductors.
  The carriers near the top of the band, having wavelength of order of a lattice spacing, undergo of order
  $10^8$ changes in the sign of their phase from one end to the other of a macroscopic sample of size $1cm$. It is
  intuitively clear that maintaining phase coherence of such a rapidly oscillating wave is impossible. As the
  antibonding electrons at the top of the band condense into the $k\sim 0$ state at the bottom of the band
  their wavelength becomes the size of the sample and their phase maintains the same sign across the
  macroscopic sample dimension, thus allowing for the existence of phase coherence over
  macroscopic length scales which is the hallmark of superconductivity.
  
  Finally if we interpret $\lambda\sim k_F^{-1}$ as the size of the electronic wavefunction, the fact that it expands and reaches the
  boundaries of the macroscopic sample in the transition to superconductivity suggests that negative charge will
  flow from the interior towards the boundaries of the sample as the normal metal becomes superconducting\cite{charge}.

 \section{Historical precedents}

There were many attempts to understand superconductivity before BCS theory. Some of these early attempts 
focused on physics closely related to what we discuss in this paper.

\noindent {\it(1)} The idea that superconductivity would occur only when the normal state carriers are holes, i.e. when the band
is almost full, was discussed in early theoretical work by Papapetrou\cite{papa}.
He argued that electrons at the top of the Fermi distribution would become metastable if the Fermi level
was close to a zone boundary. That the band should be almost full was also deemed to be essential
in the theory of Born and Cheng\cite{born}.

\noindent {\it(2)} The idea that the superconducting electrons are not sensitive to the discrete ionic potential was discussed
by Kronig\cite{kronig}. He proposed that  electron-electron interaction effects would dominate over 
electron-ion effects, and that the ionic lattice should be replaced by a continuum positive background
for the description of superconductivity. In the review by Smith and Wilhelm\cite{smith} it is also stated that the
superconducting electrons, in order to move freely, may experience
'some binding with the lattice as a whole rather than with particular atoms'. Note how different this is to BCS theory,
where coupling of electrons not only to the discrete ions but even to their deviation from equilibrium position is deemed essential.

\noindent {\it(3)} Brillouin\cite{bri} postulated that the energy versus $k$ relation in superconductors may show secondary
minima {\it near the Brillouin zone boundary} (in a band where the minimum is at the zone center), 
and that electrons in those states would not be sensitive to
scattering. 

\noindent {\it(4)} Schafroth\cite{schaf} proposed that electrons at the top of the Fermi distribution would pair into a resonant state of
$negative$ binding energy, however such that their combined energy would be less than twice the
Fermi energy of single particles, so they would not be able to break up into single particles due to
the Pauli principle. This idea obviously requires that the Fermi level be high in the band, or at the
very least not near the bottom of the band. It also foreshadows the concept of 'kinetic energy driven pairing'.

\noindent {\it(5)} Bardeen in early work \cite{bardeen} suggested that superconducting electrons would have a much smaller
effective mass than normal electrons. However he abandoned this concept   in BCS theory.

\noindent {\it(6)}  Meissner wondered whether superconductivity is carried by the same electrons that carry the
normal state current or by different ones\cite{meiss}. He favored the latter alternative, based on the
observation that atoms with only one valence electron outside a closed shell do not form
superconductors. This is in agreement with the ideas discussed here, since the antibonding
electrons do not carry electric current in the normal state, in fact they do precisely the
opposite.

\noindent {\it(7)} London\cite{london0} pointed out that diamagnetism could be understood if electrons in superconductors behave as
electrons in giant atoms. A natural extension of London's idea is that the charge distribution in superconductors will also be
inhomeogeneous as in real atoms, with more positive charge near the center and more negative charge near
the boundaries\cite{atom,charge}.

\section{The cuprates, conventional superconductors, and the pairing mechanism}

It is generally agreed that 'conventional' superconductors are described by BCS-electron-phonon theory, and that an unconventional
mechanism applies to the cuprates. There is no general agreement on which is the right mechanism for the cuprates, with proposals ranging from 
purely electronic to magnetic to electron-phonon interactions of unconventional type\cite{mech,mech2}. 
However the considerations in this paper should apply to all superconductors, conventional or otherwise,
 because they relate to fundamental aspects
of the band theory of solids. 

The predominance of hole carriers in the normal state of conventional and unconventional superconductors has been pointed out
repeatedly elsewhere\cite{papa,born, kikoin,chapnik,correl}. For the cuprates, we have proposed that
the hole carriers of interest are  
those at the top of the band formed by overlap of planar oxygen $p\pi$ orbitals pointing in direction perpendicular
to the Cu-O bond\cite{twoband}. For $MgB_2$, the holes in the nearly full bands formed by overlap of
planar boron $p\sigma$ orbitals are generally believed to be the dominant carriers driving superconductivity, and there are
also electron-like carriers at the Fermi level from other bands\cite{kortus}. 
Calculations for a two-band model, one with hole-like and one with electron-like carriers at the Fermi energy\cite{twoband},
yield results for tunneling characteristics that resemble experimental observations in $MgB_2$, with hole (electron) carriers
giving rise to a large (small) gap\cite{gonnelli}.
For
electron-doped cuprates, the existence of hole carriers in the regime where they become superconducting has been 
established experimentally\cite{electron}.

We do not address here  the question  whether   specific mechanisms  unrelated to the physics discussed here 
play or don't play a role in different classes of materials.  However  the following two points necessitate discussion:
first, is it $possible$ that 'undressing' is the driving force for superconductivity in any or in all materials?
We have shown elsewhere that within a class of models (dynamic Hubbard models) pairing leads to lowering
of kinetic energy and that as a consequence the low temperature phase is superconducting, in the absence of electron-phonon
interactions\cite{dynh,dynh2}. At the same time in these models pairing gives rise to increased quasiparticle weight and transfer of
optical spectral weight from high to low frequencies\cite{undr}. Hence at least in these models pairing and superconductivity may be
said to be 'driven' by quasiparticle undressing. Instead,  in other models
with different pairing mechanisms 'undressing' may be a $consequence$ of the transition to 
superconductivity.  The physics that is reflected in dynamic Hubbard models is not specific to one
class of materials but is generic to electrons and ions in solids as discussed in \cite{hole2}.

Second, the electron-phonon interaction is known to lead to an isotope effect in $T_c$ in most conventional materials\cite{allen}
and to an isotope effect on the London penetration depth in the cuprates\cite{isotope}.
How can this be related to the physics discussed here? First it is clear that the electron-phonon interaction
generally will modify electronic-related properties, e.g. bandgaps, due to ionic zero-point motion\cite{allen2}.
For dynamic Hubbard models we have shown that the ionic zero point motion leads to enhancement of the
correlated hopping term in the Hamiltonian and as a consequence to a positive isotope shift in $T_c$\cite{mgb2}.
 Now the London penetration depth can be written as
 \beq
 \frac{1}{\lambda_L^2}=\frac{4\pi n_s e^2}{m^* c^2}
 \eeq
 Note that only the combination $n_s/m^*$ enters this expression, with $n_s$ the density of superfluid carriers
 and $m^*$ the superfluid carrier's effective mass.
 Within the dynamic Hubbard model Hamiltonian, a reduction in $\lambda_L$ would be expected due to 
 lowering of the pair effective mass caused by ionic zero point motion\cite{mgb2,londonlength}. This may appear to be 
 incompatible with the argument in this paper that  
  the superfluid carriers completely undress from the electron-ion interaction and respond to perturbations as if
 they had the free electron mass. However the two points of view can be reconciled if 
 one interprets the superfluid weight $n_s/m^*$ as $n_s^{eff}/m_e$ and adscribes its enhancement by larger ionic zero-point motion to an enhancement of the effective superfluid density $n_s^{eff}$.
 
Table I summarizes the different properties of electrons at the Fermi energy when the Fermi level is near the bottom and
near the top of the band and some resulting properties of the solid. Within conventional BCS-electron-phonon theory, these properties do not play an important 
role in superconductivity, and superconductivity can occur with either bonding or antibonding electrons
 at the Fermi energy. Instead, 
independent of what role the electron-phonon interaction may play in superconductivity we
propose that only when a solid has at least some carriers with the characteristics listed in the right column can
superconductivity occur, and that when it does the normal state carriers of the right column adopt
characteristics of the carriers in the left column. In simple and noble metals and any other metal where
only carriers  of the
type described by the left column exist at the Fermi energy superconductivity would not occur according to our
theory, no matter how strong the electron-phonon interaction.

\begin{table}[h]
\caption{Different properties of the carriers at the Fermi energy when the Fermi level is near the bottom
(bonding electron) and near the top of the band (antibonding electron). }
{\begin{tabular}{@{}ll@{}} \toprule
{\bf Bonding electron} & {\bf Antibonding electron}  \\
{\bf at the Fermi energy} & {\bf at the Fermi energy}  \\
 \colrule
Undressed&Dressed\\
Low kinetic energy&High kinetic energy\\
Long wavelength&Short wavelength\\
Small effective mass&Large effective mass\\
Uniform charge density&Nonuniform charge density\\
Moves in direction of force&Moves opposite to force\\
Conducts electricity&Anticonducts electricity\\
Contributes to Drude weight&Anticontributes to Drude weight \\
Detached from lattice&Transfers momentum to lattice\\
Large quasiparticle weight& Small quasiparticle weight \\
Coherent conduction&Incoherent conduction \\
Large Drude weight&Small Drude weight \\
Negative Hall coefficient&Positive Hall coefficient \\
Good metals& Bad metals \\ 
Stable lattices& Unstable lattices \\ 
Ions attract each other& Ions repel each other \\ 
Carriers repel each other& Carriers attract each other \\ 
Normal metals& Superconductors  \\ 
\botrule
\end{tabular}}
\end{table}

\section{Discussion}

In this paper we have continued our analysis of the differences between electrons and holes in  
energy bands and its implications for the understanding of superconductivity. Our earlier work\cite{hole2,hole2p,dynh,dynh2,undr}
centered on the differences between electrons and holes arising from the electron-electron interaction.
Here we have focused on the even more basic aspects of electron-hole asymmetry that arise from
the electron-ion interaction.

It is interesting that the effects of electron-electron interaction and electron-ion interaction related to
electron-hole asymmetry are qualitatively similar. Both effects lead to spectral weight transfer from
low frequencies to high frequencies as the carriers evolve from electron-like to hole-like as the
Fermi level rises in the band. Both effects lead to a decrease in the electrical conductivity per
carrier as the Fermi level rises: electron-electron interactions because the carriers become
heavier, and electron-ion interaction because Bragg scattering causes the antibonding
electrons to move in direction opposite to the applied force. Fundamentally, both effects
lead to 
'dressing' of the quasiparticle as the Fermi level rises in the band, where we define 'dressing'
loosely as what makes the quasiparticle different from the bare particle, the free electron.

If dressing impairs the electrical conductivity, and if superconductors are perfect conductors of
electricity, it is natural to conclude that superconductivity has to be associated with 'undressing'.
The fact that pairing of hole carriers effectively shifts the Fermi level to a region lower in the band
where the carriers are less dressed supports this point of view. Furthermore it is natural to conclude
in view of these considerations that 'undressing' will affect both the dressing originating
in the electron-electron interaction and that originating in the electron-ion interaction.
Experiments support this interpretation.

What is however not obvious is that carriers will undress $completely$ when the transition to
superconductivity takes place, and respond as if they had the bare mass and the bare charge of the free
electron, as the experimental evidence indicates. The dynamic Hubbard models\cite{dynh}
as well as the Hubbard model with correlated hopping\cite{corrhop} predict that the
hopping amplitude increases upon pairing, hence the effective mass decreases, but they 
$do$ $not$ predict that the effective mass becomes the free electron mass. Furthermore the magnitude
of effective mass decrease predicted by the models
depends on parameters in the models and on band filling.  

The superfluid electrons in the superconducting state have a wavefunction that extends 
coherently over the macroscopic dimensions of the sample. As a consequence they no longer
'see' the discrete nature of the electron-ion potential that varies over microscopic scales, instead they see
an average smooth background of positive charge. In other words, the wavefunction is
smooth over interatomic distance scales: the carriers have undressed from the electron-ion
potential and they can no longer transfer momentum to the ionic lattice.
For this scenario to be possible, electrons at the Fermi energy have to pair, as it is the pairing
that gives rise to superconductivity and to undressing and allows the electrons to circumvent the Pauli principle. 
Beyond the pairing correlations, superfluid electrons will resemble free electrons in a smooth
positive background, a 'Thomson atom'\cite{atom}. 

The point of view discussed here also highlights the important role of the electronic confinement of electrons 
near the top of the band in the normal state, which raises their kinetic energy, and of their deconfinement upon the transition to the
 superconducting state, which lowers their kinetic energy. This is especially clear when one considers the lowest band in a solid within the weak electron-ion approximation,
but should also apply more generally. The wavelength of electrons at the bottom of the band is macroscopic and becomes gradually smaller
as the Fermi level rises, finally being of the order of the lattice spacing for the Fermi level near the top of the band. The wavefunction deconfines and the
wavelength goes from microscopic to macroscopic as the antibonding electrons at the Fermi energy condense into the
$k\sim0$ superconducting state. In metals where the wavelength of electrons at the Fermi energy is large in the normal state 
(bonding electrons), no tendency to superconductivity will exist.

In the conventional BCS-Fermi liquid theory quasiparticles are fixed objects that develop special correlations
when the transition to superconductivity occurs but do not change their intrinsic nature. Our previous work on 'undressing' from the
electron-electron interaction instead had proposed that quasiparticles do change intrinsic properties, their quasiparticle weight and the magnitude of their
effective mass, when the  transition to superconductivity occurs\cite{undr}. Here we have argued that this change in intrinsic properties is even more radical:
quasiparticles also change the $sign$ of their effective mass from negative to positive and their wavelength from
microscopic to macroscopic, when they condense into the superconducting state. 

Note that in ordinary Bose condensation for point-like bosons the phase transition as function of increasing density
occurs when the interparticle distance becomes comparable to the boson de Broglie wavelength.  Analogously here, the onset of superconductivity as function of increasing band occupation  occurs when the
Fermi level is high enough in the band such that the de Broglie wavelength of electrons at the Fermi level becomes comparable to the $interatomic$ distance. As
in ordinary Bose condensation, the transition is to a state with macroscopic de Broglie wavelength\cite{blatt}. 

In ordinary metals, charge inhomogeneity  occurs at the level of a single unit cell. If the 
superfluid electrons do not 'see' the discrete ionic lattice, the unit cell becomes the entire sample and consequently charge
inhomogeneity can occur at a macroscopic level in superconductors. Just as in the metallic unit cell bonding electrons lower their
kinetic enery by expanding their wavefunction from one atom to its neighbor, in the superconductor to lower the kinetic energy the electronic wavefunction will expand towards the outer boundaries of the sample. 
Indeed, as discussed in other work\cite{atom,charge,charge0} we
expect superfluid electrons to  have a tendency to go near the surface of the sample,
giving rise to an excess of negative charge in that region and  to experimentally observable consequences\cite{ellipsoid,electrodyn}. Furthermore we suggest that
 this expansion of
the electronic wavefunction to the boundary of the 'macroscopic unit cell' and beyond
is likely to be relevant to the understanding of the superconducting proximity effect.


\begin{references} 
\bibitem{hole2} J.E. Hirsch, Phys.Rev. B{\bf 65}, 184502 (2002).
\bibitem{hole2p}  J.E. Hirsch, Phys.Lett. A {\bf 134}, 451 (1989); Chem.Phys.Lett. {\bf 171}, 161 (1990); Phys.Rev. B{\bf 48}, 3327 (1993); J. E. Hirsch and F. Marsiglio, Phys.Rev. B {\bf 41}, 2049 (1990). 
\bibitem{dynh} J.E. Hirsch, Phys. Rev. Lett. {\bf 87}, 206402 (2001); 
Phys.Rev. B {\bf 65}, 214510 (2002); Phys.Rev. B {\bf 66},  064507 (2002);
Phys.Rev. B {\bf 67},  035103 (2003).
\bibitem{dynh2}  F. Marsiglio, R. Teshima and J. E. Hirsch,  Phys.Rev. B {\bf 68} 224507 (2003).
\bibitem{undr} J.E. Hirsch, Physica C {\bf 201}, 347 (1992); Physica C {\bf 364-365}, 37 (2001);
Phys.Rev. B {\bf 62}, 14487 (2000), Phys.Rev. B {\bf 62}, 14498 (2000).
\bibitem{atom} J.E. Hirsch, Phys.Lett. A{\bf 309}, 457 (2003).
\bibitem{corrhop} J. E. Hirsch and F. Marsiglio, Phys.Rev. B {\bf 39}, 11515 (1989).
\bibitem{matthias} The connection between lattice instabilities and superconductivity
 was repeatedly emphasized by B. Matthias, who wrote in 1973: {\it
From now on, I shall look for systems that should exist, but won't - unless one can persuade them}.
B.T. Matthias, Physica {\bf 69}, 54 (1973).
\bibitem{kittel} C. Kittel, ``Introduction to Solid State Physics'', 7th Edition, Wiley, New York, 1996.
\bibitem{harrison} W.A. Harrison, ``Pseudopotentials in the theory of metals'', Benjamin, New York, 1966.
\bibitem{peierls} R.E. Peierls, Phys. Z. {\bf 30}, 273 (1929).
\bibitem{ding} H. Ding, J. R. Engelbrecht, Z. Wang, J. C. Campuzano, S.C. Wang, 
H.B. Yang, R. Rogan, T. Takahashi, K. Kadowaki, and D. G. Hinks,
 Phys. Rev. Lett. {\bf 87}, 227001 (2001).
\bibitem{marel} H. J. A. Molegraaf, C. Presura, D. van der Marel, P. H. Kes, and M. Li
Science {\bf 295}, 2239 (2002);  A.F. Santander-Syro, R.P.S.M. Lobo, N. Bontemps, Z. Konstantinovic, 
Z.Z. Li and H. Raffy, Europhys.Lett.62, 568 (2003).
\bibitem{kikoin} I.Kikoin and B. Lasarev, Nature {\bf 129}, 57 (1932); ZhETF {\bf 3}, 44 (1933); Physik.Zeits. d. Sowjetunion {\bf 3}, 351 (1933).
\bibitem{chapnik} I.M. Chapnik, Sov,Phys. Doklady {\bf 6}, 988 (1962).
 \bibitem{sign} M. Galffy and E. Zirngiebl, Sol.St.Comm. {\bf 68}, 929 (1988); S.J. Hagen, C. J. Lobb, R. L. Greene, 
 M. G. Forrester and J. H. Kang, Phys.Rev.B {\bf 41}, 11630 (1990);
 C.C. Almasan, S. H. Han, K. Yoshiara, M. Buchgeister, D. A. Gajewski, L. M. Paulius, J. Herrmann,  M. B. Maple, 
 A. P. Paulikas, Chun Gu, and B. W. Veal, Phys.Rev.B {\bf 51}, 3981 (1995).
  \bibitem{sign2}  H. Van Beelen, JP Van Braam Houckgeest, HM Thomas,
C. Stolk, and R. De Bruyn Ouboter, Physica {\bf 36}, 241 (1967);
  C. H. Weijsenfeld, Phys.Lett. A {\bf 28}. 362 (1968);
  N. Usui. T. Ogasawara, K. Yasukochi, and S. Tomoda, 
  J. Phys. Soc. Japan {\bf 27}, 574 (1969); K. Noto, S. Shinzawa and Y. Muto, Sol.St.Comm. {\bf 18}, 1081 (1975).
 \bibitem{london} F. London, ``Superfluids'', John Wiley \& Sons, Inc., New York, 1950, Vol. 1.
\bibitem{adkins} C.J. Adkins and J.R. Waldram, Phys.Rev.Lett. {\bf 21}, 76 (1968).
\bibitem{bok} J. Bok and J. Klein, Phys.Rev.Lett. {\bf 20}, 660 (1968).
\bibitem{morris} T.D. Morris and J.B. Brown, Physica {\bf 55}, 760 (1971).
\bibitem{chiang} Y.N. Chiang and O.G. Shevchenko, Low Temp. Phys. {\bf 22}, 513 (1996).
\bibitem{becker} R. Becker, F. Sauter and C. Heller, Z. Physik {\bf 85}, 772 (1933).
\bibitem{hildebrand} A.F. Hildebrand , Phys.Rev.Lett. {\bf 8}, 190 (1964);
N.F. Brickman, Phys.Rev. {\bf 184}, 460 (1969); J. Tate, S.B. Felch and B. Cabrera,
Phys.Rev. B{\bf 42}, 7885  (1990);
A.A. Verheijen, J.M. van Ruitenbeek, R. de Bruyn Ouboter, and L.J. de Jongh,
  Physica B {\bf 165-166}, 1181 (1990); Nature {\bf 345}, 418 (1990);
M.A. Sanzari, H.L. Cui and F. Karwacki, Appl. Phys. Lett.  {\bf 68}, 3802 (1996).
\bibitem{gyro} I.K. Kikoin and S.W. Gubar, J.Phys. USSR {\bf 3}, 333 (1940); 
  R.H. Pry. A.L. Lathrop and W.V. Houston, Phys.Rev. {\bf 86}, 905 (1952);
  R. Doll, Zeits. f. Physik {\bf 153}, 207 (1958).
\bibitem{science} J.E. Hirsch, Science {\bf 295}, 2226 (2002).
\bibitem{gins} W.L. Ginsburg, Fort. Phys. {\bf 1}, 88 (1953).
\bibitem{charge} J.E. Hirsch, Phys.Rev.B {\bf 68}, 184502 (2003).
\bibitem{papa}A. Papapetrou, Z. F. Physik {\bf 92}, 513 (1934).
\bibitem{born}  M. Born and K.C. Cheng, Nature {\bf 161}, 1017 (1948).
\bibitem{kronig} R. de L. Kronig, Z. Phys. {\bf 78}, 744 (1932); {\bf 80}, 203 (1933).
\bibitem{smith} H.G. Smith and J.O. Wilhelm, Rev.Mod.Phys. {\bf 7}, 237 (1935).
\bibitem{bri} L. Brillouin, J. de Phys. et le Rad. {\bf 4}, 677 (1933).
\bibitem{schaf} M.R. Schafroth, Phys.Rev.  {\bf 96}, 1442 (1954).
\bibitem{bardeen} J. Bardeen, Phys.Rev. {\bf 81}, 829 (1951).
\bibitem{meiss} W. Meissner, Ergeb.Exakte Naturw. {\bf 11}, 219 (1932).
\bibitem{london0} F. London, Nature {\bf 140}, 793 (1937).
\bibitem{mech} P. Brusov, ``Mechanisms of High Temperature Superconductivity'', Rostov State University Publishing, Rostov on Don, 1999.
\bibitem{mech2} V. Z. Kresin, H. Morawitz and S. A. Wolf,  ``Mechanisms of Conventional and High Tc Superconductivity'', 
Oxford University Press, New York, 1993.
\bibitem{correl} J.E.  Hirsch, Phys.Rev. B {\bf 55}, 9907 (1997)
\bibitem{twoband} J. E. Hirsch and S. Tang, Sol.St.Comm.  {\bf 69}, 987 (1989); J. E. Hirsch and F. Marsiglio, Phys.Rev. B {\bf 43}, 424 (1991).
\bibitem{kortus} J. Kortus, I. I. Mazin, K. D. Belashchenko, V. P. Antropov, and L. L. Boyer, Phys.Rev.Lett. {\bf 86}, 4656 (2001)
\bibitem{gonnelli}
R. S. Gonnelli, D. Daghero,  G. A. Ummarino, V. A. Stepanov , J. Jun, S. M. Kazakov, and J. Karpinski, Phys.Rev.Lett. {\bf 89}, 247004 (2002).
\bibitem{electron} Wu Jiang, S. N. Mao, X. X. Xi, Xiuguang Jiang, J. L. Peng, T. Venkatesan, C. J. Lobb, and R. L. Greene,
Phys.Rev.Lett. 73, 1291 (1994); M. Suzuki , S. Kubo, H. Ishiguro and K. Haruna, 
Phys.Rev. B 50, 9434 (1994); Z.Z. Wang, T. R. Chien,  N. P. Ong, J. M. Tarascon and E. Wang, 
 Phys.Rev. B 43, 3020 (1991); 
 M. A. Crusellas, J. Fontcuberta and S. Pi–olM. Cagigal and J. L. Vicent, Physica C 210, 221 (1993);
 P. Fournier, X. Jiang, W. Jiang, S. N. Mao, T. Venkatesan, C. J. Lobb, and R. L. Greene, Phys.Rev. B 56, 14149 (1997).
\bibitem{allen} P.B. Allen, Nature {\bf 335}, 396 (1988).
\bibitem{isotope} GuoÐmeng Zhao, K. K. Singh, A. P. B. Sinha, and D. E. Morris,
Phys.Rev.B {\bf 52}, 6840 (1995); J. Hofer, K. Conder, T. Sasagawa, Guo-meng Zhao, M. Willemin, H. Keller and K. Kishio, 
Phys.Rev.Lett. {\bf 84}, 4192 (2000).
\bibitem{allen2} P.B. Allen, Phil. Mag. B{\bf 70}, 527 (1994).
\bibitem{mgb2} J.E. Hirsch, Phys.Lett. A {\bf 282}, 392 (2001), Sect. 5.
\bibitem{londonlength}  J. E. Hirsch and F. Marsiglio, Phys.Rev. B {\bf 45}, 4807 (1992).
\bibitem{blatt} J.M. Blatt, {\it Theory of Superconductivity}, Academic Press, New York, 1964.
\bibitem{charge0} J.E. Hirsch, Phys.Lett. A{\bf 281}, 44 (2001).
\bibitem{ellipsoid} J.E. Hirsch, Phys.Rev.Lett. {\bf 92}, 016402 (2004).
\bibitem{electrodyn} J.E. Hirsch, Phys.Rev. B{\bf 69}, 214515  (2004).



\end{references}
 \end{document}